\documentclass[twocolumn]{aastex631}
\usepackage{gensymb}

\newcommand{\Fig}[1]{Figure~\ref{#1}}

\newcommand{\Sec}[1]{Section~\ref{#1}}
\newcommand{\Secs}[2]{Sections~\ref{#1} and \ref{#2}}
\newcommand{\Eq}[1]{Equation~\ref{#1}}
\newcommand{\Table}[1]{Table~\ref{#1}}
\newcommand{\bmax}{$B_{\rm max}$}
\newcommand{\bmean}{$B_{\rm mean}$}

\usepackage{soul}
\setstcolor{red}
\newcommand{\abs}[1]{#1}

\begin{document}

\title{Analysis of BMR tilt from AutoTAB catalog: Hinting towards the thin flux tube model?}

\author[0000-0001-7036-2902]{Anu Sreedevi}
\affiliation{Department of Physics, Indian Institute of Technology (Banaras Hindu University), Varanasi 221005, India}
\email{anubsreedevi.rs.phy20@itbhu.ac.in}

\author[0000-0003-3191-4625]{Bibhuti Kumar Jha}
\affiliation{Southwest Research Institute, Boulder, CO 80302, USA}
\affiliation{Aryabhatta Research Institute of Observational Sciences, Nainital 263002, Uttarakhand, India}

\author[0000-0002-8883-3562]{Bidya Binay Karak}
\affiliation{Department of Physics, Indian Institute of Technology (Banaras Hindu University), Varanasi 221005, India}
\email{karak.phy@iitbhu.ac.in}

\author[0000-0003-4653-6823]{Dipankar Banerjee}
\affiliation{Aryabhatta Research Institute of Observational Sciences, Nainital 263002, Uttarakhand, India}
\affiliation{Indian Institute of Astrophysics, Koramangala, Bangalore 560034, India}
\affiliation{Center of Excellence in Space Sciences India, IISER Kolkata, Mohanpur 741246, West Bengal, India}

\begin{abstract}

%\abs{One of the intriguing mechanisms of the Sun in its convection zone is the development of bipolar magnetic regions (BMRs), which are the regions of concentrated magnetic fields and emerge with opposite polarities on the solar surface, as seen in LOS magnetogram observations.} 
%\abs{}
One of the intriguing mechanisms of the Sun is the formation of the bipolar magnetic regions (BMRs) in the solar convection zone which are observed as regions of concentrated magnetic fields of opposite polarity on photosphere.
%original: One of the intriguing mechanisms of the Sun in its convection zone is the formation of bipolar magnetic regions (BMRs), which are the regions of concentrated magnetic fields of opposite polarity.
These BMRs are tilted with respect to the equatorial line, which statistically increases with latitude. 
The thin flux tube model, employing the rise of magnetically buoyant flux loops and their twist by Coriolis force, is a popular paradigm for explaining the formation of tilted BMRs. 
In this study, we assess the validity of the thin flux tube model by analyzing the tracked BMR data obtained through the Automatic Tracking Algorithm for BMRs (AutoTAB). 
Our observations reveal that the tracked BMRs exhibit the expected collective behaviors.
We find that the polarity separation of BMRs increases over their normalized lifetime, supporting the assumption of a rising flux tube from the CZ. Moreover, we observe an increasing trend of the tilt with the flux of the BMR, suggesting that rising flux tubes associated with lower flux regions are primarily influenced by drag force and Coriolis force, while in higher flux regions, magnetic buoyancy dominates. Furthermore, we observe Joy's law dependence for emerging BMRs from their first detection, indicating that at least a portion of the tilt observed in BMRs can be attributed to the Coriolis force. 
Notably, lower flux regions exhibit a higher amount of fluctuations 
associated with their tilt measurement 
%BBK: Please think carefully "associated with their tilt measurement". Is the fluctuations assciated measurments which could be due to artifacts. I mean the flucuations are real. So does this not confuse the reader? Think.
compared to stronger flux regions, suggesting that lower flux regions are more susceptible to turbulent convection.

\end{abstract}


\section{Introduction} \label{sec:intro}

The Bipolar Magnetic Regions (BMRs), commonly observed in line-of-sight (LOS) magnetograms, emerge on the solar surface in an East-West orientation with a 
finite 
tilt having the leading polarity closer to the equator \citep{HE1919}. 
Statistically, BMR tilt  
increases with the latitude of 
emergence, which is 
commonly referred to as Joy's law.
This behavior was validated by 
many authors, 
including, \citet[]{WS1991, H1991, SG1999}  
using white-light observations prior to the availability of magnetograms. The tilt in the BMRs plays an essential role in the reversal of the existing poloidal field through the dispersal and cancellation of fluxes on the solar surface \citep[]{B1961, L1964}. This process is popularly known as the Babcock--Leighton mechanism and is an essential component of solar dynamo \citep[e.g.,][]{CS2023, K2023}.

Our present understanding attributes the formation of BMRs to the magnetically buoyant, large-scale toroidal flux tubes generated by the solar dynamo process at the base of the convection zone \citep[BCZ;][]{P1955}. Numerical simulation studies, assuming the thin flux tube approximation, have focused on understanding the dynamics of $\Omega$-shaped flux tubes and provide constraints to the magnetic field strength in the CZ \citep{CM1995, F2009}. The works of \citet{DC1993} and \citet{FF1994} demonstrate
that the tilt in the BMR is due to the action of Coriolis force on the diverging flows at the apex of rising flux tubes. Consequently, stronger BMRs, with higher magnetic field strength at the BCZ, are expected to ascend rapidly through the CZ and experience Coriolis force for less time, leading to a reduced amount of tilt in them. 
Their works also predicted that  
with the increase of the flux in the tube, the BMR tilt increases due to the effect of the drag.
Observational studies, such as as by \citet{TL2003} using magnetic field observations from Huairou Solar Observatory Station, validate these findings, noting an initial increase in tilt angle with flux followed by a decrease for higher flux BMRs. Similar variations in tilt angle with magnetic field strength are observed in simulations by \citet{WF2011} and in observational studies by \citet{JK2020}. Interestingly, 
\citet{SK2012} did not find any systematic dependence of tilt on flux from LOS magnetic field observations of the Michelson Doppler Imager (MDI).

The theory of Coriolis force as the reason behind tilted BMR is very promising and has been studied extensively. If this theory is true, then BMRs should emerge on the photosphere with a definite tilt. 
Analyzing 715 BMRs from MDI magnetograms during 1996--2008,  \citet{SK2008} found Joy's law behavior of BMR tilts at their mid-emergence phase, concluding, `The observations indeed show the predicted latitudinal dependence (Joy’s law) and indicate that the tilt is formed below the surface.' 
Later, \citet{SB2020} noted that BMRs emerge with zero tilt and develop tilt in accordance with Joy's law at a later stage in their lifespan, challenging the thin flux tube model's prediction. They suggested that the observed Joy's law behavior is because of the inherent north-south separation of BMRs as they reach the surface. 
 %Moreover, \citet{SK2008} noted the statistically significant Joy's law trend during the mid-phase of BMR evolution in agreement with the predicted latitudinal dependence (Joy’s law) and concluded that the tilt is formed below the surface itself.
 %BBK: Bibhuti, I understand that you chagnged above sentence, however, I feel to keep the exact statement what was made in their paper because in their later paper they do not support thin flube model based on the flux dependence. Also I moved this setence up.

Joy's law is statistical, and thus, the latitudinal dependency of tilt is evident after averaging over large data samples. Persistent, significant scatter around Joy's law is consistently observed in various observational studies across different data sets \citep[e.g.,][]{H1996, WS1991, DS2010, MN2013}.  
\citet{FF1994} and  \citet{LF1995} hint towards the role of turbulent convection on rising flux tubes as the possible cause of the scatter in tilt angle. 
 This observed behaviour was further supported by the simulation of \citet{WF2013}. 
 Recently, a thorough analysis on the inconsistencies in Joy's law has been explored  by \citet{WN2023}.

Currently, observational support for the thin flux tube model is very limited and not adequate enough to establish their existence. 
Lack of dependence of BMR's tilt on flux has been used to strongly rule out the thin flux model based on studies presented in \citet{SK2008, SB2019}. While the latter study was based on a limited data set, the study of \citet{SK2012} did not include the tracked information of BMRs and hence they counted every detection of BMR as a new one, giving a higher weightage to long-living BMRs (bigger ones). 
On the basis of all these results, we are still not in a position to completely rule out the thin flux tube model.

This manuscript aims to understand the origin mechanism of BMRs using the tracked BMR information from the Automated Tracking Algorithm for BMRs \citep[AutoTAB;][]{JP2021, SJ2023} catalog, which contains the tracked information of 12,610 unique BMRs for the period 1996\,--\,2023. We study the general behavior of BMR properties over their lifetime and explore the validity of the thin flux tube model. Before we present our study, in \Sec{sec:expect}, we list what kind of observational signatures we expect based on this theory. 
%In the later section, we also try to understand the physics behind the inherent scatter in Joy's law. 
Then in \Sec{sec:results} we present our result followed by conclusions in \Sec{sec:conclusion}.

\section{Observational expectations of the thin flux tube model}
\label{sec:expect}

The simplest explanation for the formation of the BMR is given by the thin flux tube model \citep{P1955}. The model assumes a thin, untwisted flux tube with a diameter negligible compared to the length scale of perturbation in the tube, anchored in the deep CZ. Upon becoming magnetically buoyant, tubes rise through the CZ to the photosphere. If the flux tubes remain anchored in the CZ, they rise in the form of an $\Omega$-shaped loop structure. As the flux tubes rise, the draining  
plasma 
from the apex of the tube is subjected to Coriolis force, which leads to the observed tilt in the BMRs \citep{DC1993}.

Therefore, if $\Omega$-shaped loops do exist and they are responsible for the formation of the BMRs, one can verify this idea by tracking the separation of BMR's opposite polarities during the initial phases of their evolutions. We expect the polarities to separate as BMRs mature, indicating the rise of the $\Omega$ loop when observed in LOS magnetograms.

Second, if the Coriolis force acts on the rising flux tubes, then the leading polarity of the BMR is expected to be closer to the equator, and BMR's tilt will have a latitudinal dependency (e.g., see Section~2 of \citet{DC1993} for rough calculations). In such a scenario, we anticipate the BMRs will emerge with a definite tilt, and they are expected to follow Joy's law due to the pronounced effect of the Coriolis force in the CZ.

Finally, BMRs with higher flux are expected to have higher tilt as predicted by the empirical relation given in \citet{FF1994} for the tilt angle ($\gamma$),
\begin{equation}
\label{eq:1}
\gamma \propto \mathrm{sin}\lambda B_0^{-5/4} \Phi^{1/4}.
\end{equation}
Here, $B_{0}$ is the initial magnetic field in the flux tube (forming BMR) at the BCZ, $\Phi$ is the magnetic flux inside the rising flux tubes, and $\lambda$ is the emerging latitude of the flux tube.

We note that, in the above model (\Eq{eq:1}), $B_{0}$ and the $\Phi$ are made independent of each other, while in the Sun this may not be true.
Nevertheless, based on the above relation, we expect that the tilt ($\gamma$) decreases with the increase of magnetic field strength ($B_0$). Although the magnetic flux that is measured in the BMR ($\Phi_m$) is not exactly the same as $\Phi$, they are expected to be related. However, for the magnetic field it is not obvious; field that is observed in the BMR on the solar surface is quite different from the initial field $B_{0}$ of the flux tube. Despite this, we shall also check the dependence of tilt with the measured magnetic field in addition to the dependency on the flux in the BMR.  To the best of our knowledge, no definitive observational evidence supporting the thin flux tube model was conclusively confirmed. In \Sec{sec:results}, through the catalog of AutoTAB, we evaluate if the BMR properties align with the mentioned findings from the thin flux tube model.

\section{Data and Method} \label{sec:data}

We start with a brief description of the AutoTAB catalog \citep[]{SJ_iaus1, SJ2023} analyzed in this study. The catalog encompasses tracked information of 12,173 BMRs during 1996\,--\,2023, generated using AutoTAB. These tracked BMRs are observed in the LOS magnetograms from MDI \citep[1996--2011;][full cadence]{SB1995} 
and Helioseismic and Magnetic Imager \citep[HMI: 2010--present;][96~minutes cadence]{SS2012} 
onboard the Solar and Heliospheric Observatory (SOHO) and Solar Dynamic Observatory (SDO), during the period of September 1996 to December 2023 which includes complete Cycles~23, 24, and early 25. 
The working of AutoTAB is summarized below while the detail is published in \citet{SJ2023}\footnote{The AutoTAB catalog will be publicly available along with the codes at \url{https://github.com/sreedevi-anu/AutoTAB}.}.

To automatically detect the BMRs from the LOS magnetogram, AutoTAB uses a similar method prescribed by \citet{SK2012}, which was also adopted by \citet{JK2020} with slight modifications.  
% Here, we do not restrict the detection of the BMRs to inside of 0.9~\Rsun; this is done to improve the longevity of the tracking duration.
%AB+BBK: To discuss
%\abs{}
All the detected regions satisfy the flux balance condition, similar to \citet{SK2012} and \citet{JK2020} to ensure that the amount of positive and negative flux are balanced in each BMR. This flux balance condition only checks for the amount of flux (positive and negative), rather than the distribution of flux in BMRs. Hence, the AutoTAB catalogue (this work) also includes the multi-polar regions which are very frequent and thus important in contributing to the polar field in Sun \citep{Y2020}.
%along with bipolar regions.}
The detected BMRs are saved as binary files, and AutoTAB uses these files to track the detected regions. These regions undergo  pre-processing steps before the tracking for improved tracking efficiency, followed by the technique of feature association in successive binary files to track the BMR in future instances; see Sections~2.2 and 2.3 of \citet{SJ2023} for details of AutoTAB. 
It has to be noted that the detection and tracking algorithms operate independently. Therefore, AutoTAB gives users freedom to choose a different methods of detection for BMR or any other features, which can be efficiently tracked by AutoTAB.

AutoTAB tracks BMRs through their evolution on the nearside of the Sun, and the tracked BMRs have a wide range of lifetime\footnote{The time period for which AutoTAB tracks the BMR.}, which includes those tracked from emergence to decay, those tracked in their evolutionary stages only, and a small category of BMRs that live for less than 8~Hrs (mostly the ephemeral regions). These three groups were respectively classified as ``Lifetime (LT)", ``Diskpassage (DP)", and ``Short-lived (SL)" by \citet{SJ2023}.
The BMRs falling in the latter class are excluded from this study and will be explored in detail in the forthcoming manuscript. 
Lifetime BMRs are mostly composed of moderately small BMRs within the flux range of $10^{20}$ -- $10^{22}$ Mx; some of them may not produce enough contrast in white-light images to be visible as sunspots. Meanwhile, the DP class constitutes BMRs, which have been only tracked during their appearance on the near side of the Sun. Therefore, this class includes 
(i) BMRs that are first detected near the east limb ($< 45\degree$~E) and disperse on the nearside; (ii) BMRs that emerge on the nearside of the Sun but cross the west limb ($> 45\degree$~W); and (iii) BMRs that first detected near the east limb ($<45\degree$~E) and cross the west limb ($ > 45\degree$~W). Hence, this class mainly comprises larger and stronger BMRs, exhibiting consistent evolution of magnetic properties throughout their lifetime. The majority of the tracked BMRs fall into this category \citep{SJ2023}.

AutoTAB catalog includes the tracked information of all BMRs and their physical parameters, such as maximum magnetic field (\bmax), average magnetic field (\bmean),
total unsigned flux ($\Phi_m$), 
tilt ($\gamma$), as well as positional information (latitude $\lambda$ and longitude $\phi$), at each instance during their lifetime.  The tilt of a BMR is calculated similarly to \citet{SK2012} and \citet{JK2020}, which varies between $\pm90\degree$. The convention followed here is that the BMRs, which strictly follow Joy's law, have a positive tilt in the Northern Hemisphere, whereas it is negative in the Southern Hemisphere. %\abs{}
Furthermore, the tilt of the BMRs emerging in the southern hemisphere is multiplied by a factor of $-1$ to match with the tilts of the northern hemisphere assuming hemispheric symmetry in tilt distribution. This convention is followed throughout the analysis.

\subsection{Assigning Physical Parameters to BMRs}
\begin{figure}
\centering
\includegraphics[width=0.47\textwidth]{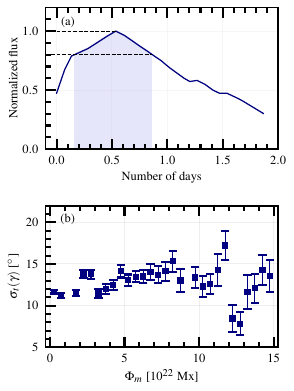}
\caption{(a) Evolution of magnetic flux ($\Phi_m$) of a representative BMR tracked by AutoTAB. The shaded area indicates the duration during which the measured $\Phi_m$ is greater than 80\% of the maximum $\Phi_m$ recorded for the BMR. (b) 
%Variation of the standard deviation of BMR tilt ($\sigma_t(\gamma)$) throughout the tracking period as a function of flux. 
Mean of the standard deviation of BMR tilt, $\sigma_t(\gamma)$ in each 5 $\times$ 10$^{21}$~Mx flux bin plotted against $\Phi_m$ of the DP class BMRs.}
\label{fig:flux_evol}
\end{figure}

In most of the previous studies, every observation of a BMR 
is treated as a new BMR. 
%In that case, it is easy to assign the tilts of BMRs or any other physical parameters. 
However, since the AutoTAB catalog provides the tracked information, we need to find a way to study the statistical properties of BMRs by assigning a unique number for BMR parameters. Thus, to get the single representative parameter for LT class BMRs, we picked the time period during which the measured $\Phi_m$ exceeds 80\% of its maximum value during tracking. Following that, we calculate the average properties only during that period.  To avoid projection effects, the maximum flux $\Phi_m$ of a BMR is considered only 
 when it resides within $55\degree$ E--W. 
In \Fig{fig:flux_evol}(a), we illustrate the evolution of flux of a typical BMR from the AutoTAB catalog. Here, shaded region represents the time window when $\Phi_m$ is more than 80\% of maximum $\Phi_m$, and thus, the physical parameters (\bmean, \bmax, $\Phi_m$, $\gamma$, and $\lambda$) are averaged over this time window.

Using this approach, we can easily assign a single value of a physical parameter for the LT class BMRs. However, assigning a single value to a BMR in the DP class, may not be appropriate as they are tracked at different evolutionary stages and we may not have the maximum $\Phi_m$ during this period. For the tilt angle, it may be more questionable to assign a single value as the fluctuations in the tilt is more pronounced compared to other parameters (refer to \Fig{fig:app2} for case studies). 

To gauge these fluctuations in tilt angles over BMR's tracking period, we calculate the standard deviation $\sigma_t(\gamma)$ in tilt for a BMR over its lifetime for all the DP class BMRs.
%To understand the dependence of scatter in tilt on the flux content $\Phi_m$ of BMRs, in 
The variation of the mean of $\sigma_t(\gamma)$ in each flux bin of $10^{22}$~Mx is presented as a function of $\Phi_m$ in 
\Fig{fig:flux_evol}(b).

%\abs{}
While the fluctuation in the measured tilt over the tracked lifetime ($\sigma_t(\gamma)$) is even larger than the mean tilt, the trend in this figure suggests a relatively stable variation with respect to the flux.  
%original: relatively stable scatter in the tilt 
We however observe a potential decrease in $\sigma_t(\gamma)$ for higher flux BMRs. The larger error bars at higher flux ranges result from a smaller number of BMRs in these bins. Hence, in this study, DP class BMRs 
can be uniformly analyzed and the same method for assigning the parameter values for the LT class can be followed for the DP class as well.
%, disregarding the phase of evolution at which AutoTAB tracks these BMRs.

 \Fig{fig:flux_evol}(b) also suggests that BMRs having higher flux are less affected by convective buffeting, possibly due to quenched convection around them and a larger amount of magnetic tension; more on this will be discussed in \Secs{JL}{sec:scatter}.

After outlining the method for assigning representative values to each tracked BMR, we proceed to assess the statistical behavior of BMRs.

\section{Results and Discussion} \label{sec:results}

\begin{figure*}
\centering
\includegraphics[width=\textwidth]{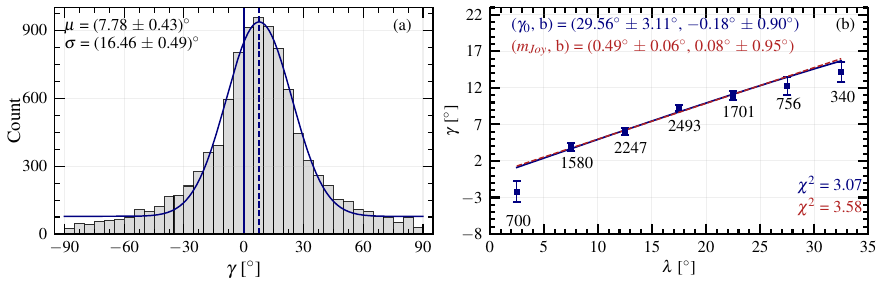}
\caption{(a) Tilt Distribution: Number of BMRs in $5\degree$ tilt bins are shown as bars. The blue solid line represents the Gaussian fitted curve (with an offset) 
with fitting parameters mentioned in the panel.
%, along with the $\chi^2$ value for the fit. 
The vertical solid blue line represents the $0\degree$ tilt, and the dashed blue line represents the Gaussian fitted mean at $7.78\degree$. (b) (Gaussian) Mean tilt in each $5\degree$ latitude bin as a function of the latitude. Blue solid and red dashed lines represent Joy’s law ($\gamma = \gamma_{0}\sin\lambda + b$) and straight line fits ($\gamma = m_{\rm Joy}\lambda + b$) with fitting parameters mentioned in the panel along with  $\chi^2$ value for the fit at the bottom right. The numbers appearing below the points display the total number of BMRs in the associated bins.}
\label{fig:fig1}
\end{figure*}

\subsection{Statistical Properties of tracked BMRs}\label{stat_prop}

We start validating the AutoTAB catalog by studying its statistical properties and comparing them with the anticipated collective behaviors of BMRs based on previous studies \citep[e.g.,][]{H1991, H1996, SG2007}. This serves as a preliminary step before investigating the observational signature for the thin flux tube model. 

\subsubsection{Distribution of tilt}
\label{tilt_dist}
One of the best-known properties of tilt is its distribution. Hence, we start by evaluating whether the assigned tilts of BMRs shows the familiar distribution,
as observed in earlier studies by \citet[]{WS1991, DS2010} and many others. In \Fig{fig:fig1}(a), we show the distribution of tilts for all the BMRs from all latitudes. %\abs{\hl{The distribution is plotted with the assumption of the symmetric hemisphere, so the tilt of the BMRs appearing in the southern hemisphere is multiplied by a factor of -1 to match with the tilts of the northern hemisphere.}} 
From this figure, 
we note that the tilt shows the well-known Gaussian distribution as reported in earlier works.
The least-square fit to the distribution with Gaussian function (shown by blue solid line in  \Fig{fig:fig1}a) gives mean ($\mu$) of $7.78\degree$ and a standard deviation ($\sigma$) of $16.46\degree$. These values closely align with those obtained by \citet[]{JK2020}, where they have not tracked BMRs and treated each detection as an independent measurement. We also compared our distribution with \citet{SK2012} by only considering BMRs in the latitude range of 15\degree\,--\,20\degree\ in both hemispheres and we find $\mu = 9.27\degree$ and $\sigma = 14.90\degree$, which show an excellent agreement.
%with \citet{SK2012}.

We further analyzed the tilt distribution 
 for individual cycles (Cycles 23 \& 24) and separately for each hemisphere and the fitting
 parameters are presented in \Table{tab:tab1}. Obtained $\mu$ and $\sigma$ values do not show a significant variation. 
Therefore, for our analyses, we combined the data from two hemispheres.

\begin{deluxetable}{lrcc}
 \tablecaption{Fitting parameters of the tilt distribution for different cycles and hemispheres.
 \label{tab:tab1}}
 \tablehead{\colhead{Cycle} & \colhead{Hem.}  & \colhead{$\mu$ (deg)}  & \colhead{$\sigma$ (deg)}} 
 \startdata
 Cycle 23  & North    & 7.72 $\pm$ 0.59  & 15.76 $\pm$ 0.67\\ 
           & South    & 7.17 $\pm$ 0.45  & 16.76 $\pm$ 0.52\\ 
           & Combined & 7.42 $\pm$ 0.42  & 16.32 $\pm$ 0.48\\
 Cycle 24  & North    & 8.52 $\pm$ 0.56  & 16.16 $\pm$ 0.63\\ 
           & South    & 7.13 $\pm$ 0.40  & 17.13 $\pm$ 0.46\\ 
           & Combined  & 7.55 $\pm$ 0.55  & 16.92 $\pm$ 0.64\\
 \enddata
\end{deluxetable}

\subsubsection{Joy's law}
\label{in_jl}
Another well-known property of BMRs is that they show a systematic dependence on latitude, i.e., Joy's law. Therefore, we plotted the mean tilts of BMRs in each 5\degree\ latitude bin by folding both hemispheres; see \Fig{fig:fig1}(b). Here, we calculate the mean by fitting a Gaussian function in the distribution of tilt in each latitude bin. 

We fit these mean values with the standard Joy's law function, i.e., $\gamma= \gamma_{0}\sin\lambda + b$, 
 (blue solid line), which yields Joy's law slope $\gamma_0$ of $29.56\degree \pm 3.11\degree$, in agreement with previous reports \citep{SK2012, JK2020}. If we force the line to pass through the origin (i.e., $b=0$), the $\gamma_0$ gets modified to $28.98\degree \pm 3.10\degree$.

Sometimes, instead of sinusoidal dependence, a linear dependence ($\gamma = m_{\rm Joy}\lambda + b$) is also used for the tilt-latitude relation \cite[e.g.,][]{WS1991, SG1999}. 
In \Fig{fig:fig1}(b), we also fit the mean $\gamma$ with this linear fit, which is shown by the dashed blue line (almost on top of red). The slope of the fitted line is 0.49 when $b\neq0$ and 0.50 when $b=0$. These values are significantly higher than the previously reported values of 0.26 and 0.28, based on white-light observations at Mount Wilson and Kodaikanal Observatories, respectively \citep{DS2010}. One reason behind this discrepancy is the difference in data type, as magnetogram tends to give a higher slope of Joy's law than the white light \citep{WC2015}. Since there is no significant difference between the fitted lines in \Fig{fig:fig1}b, we use the sinusoidal dependence as the standard Joy's law.

\subsubsection{Flux vs magnetic field}\label{flux_bmax}
\begin{figure}[t]
\centering
\includegraphics[width=0.47\textwidth]{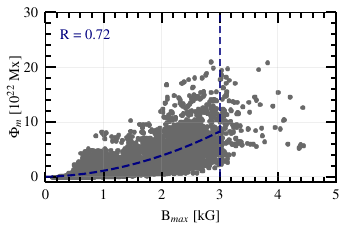}
\caption{Scatter plot between the measured \bmax\ and $\Phi_m$. The dashed blue line represents the %cubic polynomial 
quadratic fit (see text). 
%(\bmax$ = a\Phi_m^3 + b\Phi_m^2 + c\Phi_m$) with parameters $a = (0.08 \pm 0.03)$, $b = (0.56 \pm 0.1)$, and $c = (0.48 \pm 0.09)$. 
Flux values are capped at $3 \times 10^{23}$~Mx, and the outliers beyond this range attribute to the effects of defective pixels and faulty detection. }
%The outliers in the scatter plot are due to the effect of defective pixels in the data while measuring \bmax.}
\label{fig:fig5a}
\end{figure}

Assigned \bmax\ and $\Phi_m$ values of the tracking data from AutoTAB reveal a correlation between them, depicted in a scatter plot of \Fig{fig:fig5a}. The (Pearson) correlation coefficient between the quantities is 0.72, suggesting a good correlation between \bmax\ and $\Phi_m$. A similar trend (not shown) is observed between \bmean\ and $\Phi_m$ as well. It is important to note that the measured \bmax\ values are affected by the saturation limit of MDI and HMI, but the saturation effects will not significantly influence the measurement of $\Phi_m$ of the regions \citep{HL2014}. Therefore, the observed stabilization of \bmax\ with $\Phi_m$ beyond \bmax\  of 3~kG in \Fig{fig:fig5a} may reflect this phenomenon and measurements in high field regions will be significantly constrained by instrument saturation. 
\abs{The quadratic
fit is slightly better and the relation becomes \bmax\ below 3~kG is given by \bmax$ = (0.83 \pm 0.02) \Phi_m^2 + (0.27 \pm 0.04) \Phi_m$. (If we fit with cubic relation, then the equation reads \bmax$ = (0.08 \pm 0.03) \Phi_m^3 + (0.56 \pm 0.1) \Phi_m^2 + (0.48 \pm 0.09) \Phi_m$.)}
The cut-off of 3~kG is set to address the instrument saturation. Given this limitation, the relation in \Fig{fig:fig5a} suggests stronger BMRs are associated with higher magnetic flux.

\subsection{Validation of thin flux tube model}
After discussing the collective behaviors of the tracked regions, in this section, we delve into an examination of whether the tracked information from AutoTAB aligns with the expected outcomes proposed by the thin flux tube model as discussed in \Sec{sec:expect}.

\subsubsection{Evolution of footpoint separation of BMRs} \label{footpoint}
To assess the validity of the rising flux tube model, i.e., the rise of $\Omega$ loop, we calculate foot-point separation ($D$) at each time step of observation throughout the lifetime of the LT class BMRs. Here, $D$ is defined as the 
%great circle 
angular 
distance between the two polarities of the BMR at each instance of the tracking of a BMR calculated using,
\begin{equation}
\label{eq:2}
D = {\rm R_\odot} \cos^{-1}\left[\cos{\lambda_1}\cos{\lambda_2}\cos(\phi_2 - \phi_1) + \sin{\lambda_1}\sin{\lambda_2}\right].
\end{equation}

\begin{figure*}[t]
\centering
\includegraphics[width=\textwidth]{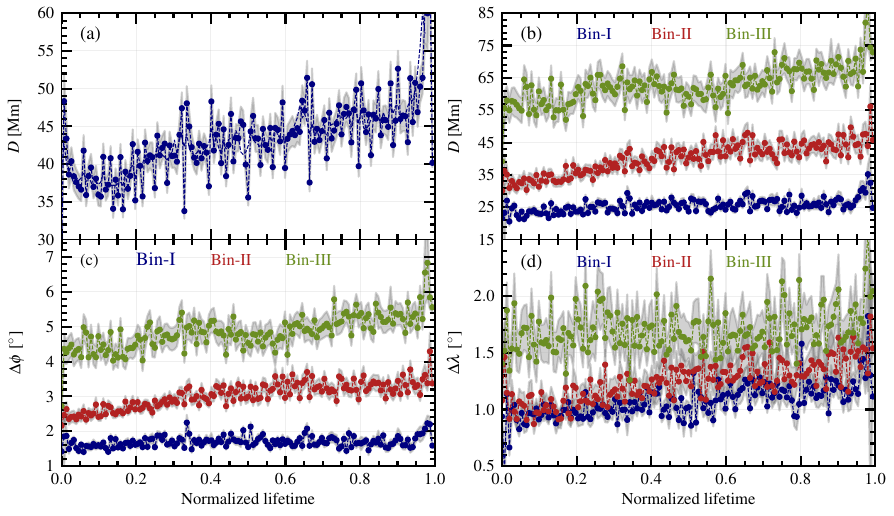}
\caption{(a) General evolutions of footpoint separation of BMR ($D$) over its lifetime.
(b) Shows the same but in three different flux ranges:  Bin-I: $10^{20}$ -- $10^{21}$~Mx (blue), Bin-II: $10^{21}$ -- $10^{22}$~Mx (red), and Bin-III: $10^{22}$ -- $10^{23}$~Mx (green). (c) and (d) show the same as in (b) but for the longitude separation ($\Delta\phi$) and latitude separation ($\Delta\lambda$), respectively.   
%, (b) $D$ 
%%over its lifetime time 
%in three different flux ranges:  Bin-I: $10^{20}$ -- $10^{21}$~Mx (blue), Bin-II: $10^{21}$ -- $10^{22}$~Mx (red), and Bin-III: $10^{22}$ -- $10^{23}$~Mx (green). (c) longitude separation ($\Delta\phi$) of BMR over its lifetime in three different flux ranges as in (b).
%%, Bin-I (blue), Bin-II (red), and Bin-III (green).  
%(d) Latitude separation ($\Delta\lambda$) of BMR over its lifetime in three different flux ranges as in (b), Bin-I (blue), Bin-II (red), and Bin-III (green). 
Note that the time is normalized to their lifetimes, and the quantities ($D$, $\Delta\phi$, and $\Delta\lambda$) are averaged over all BMRs in each 96~minute cadence. }
\label{fig:fig2}
\end{figure*}

The heliographic location of the polarities is calculated based on the flux density-weighted mean location of BMR's polarity. In \Fig{fig:fig2}(a), we show the foot-point separation ($D$) of BMR polarities as a function of time, normalized by 
%the longest living BMR (having a lifetime of 10.62~days) in the LT class. The lifetime of the longest living BMR is normalized to make it 1 unit, and others are normalized with the same normalizing factor 
%BBK: This is wrong statement. I edited it.  
their lifetimes 
to bring them on the same footing.

In \Fig{fig:fig2}(a), we see a steady, rapid increase in $D$ during the initial phase of the evolution of the BMRs, followed by a slow increase in the later stage. However, the evolution of $D$ of a typical BMR, shown in Figure~1 of \citet{SK2008},  depicts a downward trend in the footpoint separation towards the disintegration phase of the BMR, which is not evident in \Fig{fig:fig2}(a). 
To further explore the behavior observed by \citet{SK2008}, we look at a few individual BMRs (\Fig{fig:app2} in Appendix). We observe that in some cases, there is indeed a downward trend in $D$ towards the later phase of BMR life (see \Fig{fig:app_tilt}(a), (f), (g), (i), (k)). However, we find a statistically consistent increase of $D$ along with saturation in the later part of BMR's lifetime, which aligns with the expectation of the rise of $\Omega$ shaped flux tube.

 % \citet{SK2008} has previously studied the individual active region, indicating the ascent of magnetic flux tubes through the CZ. Here, we study if the tracked BMRs of AutoTAB catalog collectively, which also exhibit similar behavior.

% To evaluate this, we collect all the LT class BMRs and normalize their lifetime, with respect to the longest living BMR from the class (10.62~days), to study how collectively footpoint separation ($D$) evolves during the life of a BMR. This is presented in \Fig{fig:fig2}(a).

A closer inspection of \Fig{fig:fig2}(a) shows a rapid increase in the footpoint separation occurs during the initial phase (5\% to 30\% of lifetime).
%, suggesting the rapid ascent of flux tubes during this period. 
This accelerated growth phase in BMR evolution was previously reported as Phase-1 (acceleration) by \citet{SB2019}. The rate of rise in footpoint separation slows down in the later phase, suggesting that this can be Phase-2 (deceleration) as suggested by \citet{SB2019}. To further investigate this behavior, we segregate the BMRs into three different flux ranges based on the assigned flux, Bin-I: $10^{20}$ -- $10^{21}$~Mx, Bin-II: $10^{21}$ -- $10^{22}$~Mx, and Bin-III: $10^{22}$ -- $10^{23}$~Mx and we study the evolution of $D$ and the major contributors to $D$, i.e., $\Delta\phi$, and $\Delta\lambda$ in each of the flux bins in \Fig{fig:fig2}(b), (c), and (d), respectively.  
%The major contribution to calculating great circle distance is the difference in latitude ($\Delta\lambda$) and longitude ($\Delta\phi$) of the opposite polarities. 
\Fig{fig:fig2}(b) represents the evolution of $D$ in various flux bins. The increase in $D$ observed in \Fig{fig:fig2}(a) is primarily contributed by Bin-II, and the trend also suggests that higher flux regions are associated with higher $D$, which is consistent with previous findings \citep{SB2019}. The evolution of $\Delta\phi$ in panel~(c) reflects the rapid increase seen in $D$ in higher flux ranges (Bin-II and Bin-III); however, in \Fig{fig:fig2}(d), such a trend is not evident in the evolution of $\Delta\lambda$. %\abs{}
We also would like to point out that the trend depicted in \Fig{fig:fig2}(a) persists across the latitudinal bins, with their patterns overlapping.

It has been suggested in previous simulation studies by \citet{DC1993, SR2005} that the flux tubes get tethered from the CZ as they rise up. If the tethering happens, the reflection of the same can 
% be assumed to 
be seen as a rapid change in longitude separation and a steady change in latitude separation due to imparted circular motion from the twisted flux tubes. Now, the question arises, when does the tethering happen in a flux tube? One can assume that flux tubes with lower strength and lower flux content will tether from the CZ at an earlier time compared to stronger flux tubes. 
%Anu: Should we add the twist here? Because the thin-flux tube model assumes no twist right? But we know there'll be twist in BMRs and that can be the increase we see in latitudes

The rapid change in $\Delta\phi$ in higher flux bins of \Fig{fig:fig2}(c) may be attributed to this effect. However, we fail to find any statistically significant variation in $\Delta\lambda$ in \Fig{fig:fig2}(d). An important consideration here is that the BMRs in LOS magnetogram data are detected only once the flux tubes emerge from the photosphere radially. Additionally, AutoTAB tracks them exclusively when both polarities have emerged and they hold a strong flux balance condition. Hence, the dataset lacks information about the onset emergence phase of the BMR. The absence of a rapid increase in lower flux regions could be because tethering had already occurred before AutoTAB started tracking BMRs. This is in contrast to higher flux regions, where tethering occurs at a later stage, allowing AutoTAB to track them during the tethering period effectively and reflected as the rapid increase in $\Delta\phi$, seen in \Fig{fig:fig2}(c).

\subsubsection{Tilt angle at the first detection} 
\label{JL_initial}

The evolution of $D$ lends support to the assumption of ascending flux tubes associated with the formation of BMRs. One debated point has been whether the BMRs emerge with definite tilt or they acquire the observed tilt after emergence. 
%Hence, in this section, we validate if Coriolis force can be attributed to the observed tilt in the BMRs. 
According to the thin flux tube model, BMRs should have acquired tilt as the flux tubes rise through the CZ. Hence, we anticipate a definite tilt angle in BMRs at the emergence. To evaluate this, we collected all those BMRs that emerged on the near side of the Sun and looked for their tilt and Joy's law behavior at the very first detection. 
We emphasize that the latitude and tilt values considered here correspond to the BMR's first detection. To avoid the projection effect, we restrict the BMR emergence between $45\degree$ E-W longitudes. AutoTAB tracks 5635 such BMRs, which lie within the flux range of  $10^{19}$ -- $10^{23}$~Mx with the median at $4.05 \times 10^{21}$~Mx.

\begin{figure}%[tbh!]
\centering
\includegraphics[width=0.49\textwidth]{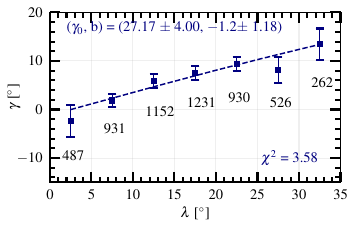}
\caption{ Joy's law plot at the first detection of BMR emerging between $45\degree$ E-W. (Gaussian) Mean tilt in each $5\degree$ latitude bin as a function of the latitude.  Blue dashed lines represent Joy’s law fit ($\gamma = \gamma_{0}\sin\lambda + b$) with parameters and $\chi^2$ value for the fit mentioned in the panel.  Numbers appearing below the points mark the total number of BMRs in the associated bins.}
\label{fig:fig3}
\end{figure}

In \Fig{fig:fig3}, first, we plot the Gaussian mean in each latitude bin of 5\degree\ and then fit Joy's law function using the least square fitting method. Here we note that the BMRs show a definite tilt at their first detection, which increases with latitude, as we expect from Joy's law. However, Joy's law amplitude of $\gamma_0=27.17\degree$   
implies a somewhat weaker latitude dependence compared to \Fig{fig:fig1}(b). 
%\red{The smaller value of $\gamma_0$ is expected as BMR emerges with a lesser tilt, and it increases in the later phase of BMR's life, which is consistent with the previous findings \citet{SK2008, SB2019}.} 
Comparing this $\gamma_0$ with that obtained during the matured phases, we find that the tilt increases in the later phase of BMR's life, which is consistent with the previous findings \citep{SK2008, SB2019}. 
This is also visible in the individual case studies in \Fig{fig:app_tilt}. 
Nevertheless, based on the behavior observed in \Fig{fig:fig3}, we can say that, statistically, BMRs emerge with a definite tilt in accordance with Joy's law.

\citet{SB2019, SB2020} suggested that BMRs emerge with zero tilt and develop tilt in accordance with Joy's law in the later part of their life from the analysis of the Solar Dynamic Observatory Helioseismic Emerging Active Region (SDO/HEAR) survey, thereby arguably ruling out the Coriolis force as a possible cause for the tilt in the BMRs. The active regions chosen for their study mainly had two criteria: (i) The regions should appear in the continuum observation, and (ii) Regions should emerge in the quiet Sun region and not in a region close to existing AR. We argue that these criteria may lead to selection bias, and the result may be affected by this.  
To validate if the BMR emerges with zero tilt or not, we conducted some case studies of BMR. 
We remind that AutoTAB tracks the BMR only for the time period in which the flux balance condition holds, and the initial developing signatures of BMRs might have been missed in tracking. 
Therefore, we select some BMRs at various stages of solar cycles and flux ranges and go back in time (time before they are detected by AutoTAB) to observe their tilt at the emergence phase.  
While making our selection, we make sure that the latitude of the emergence of these BMRs is between $\pm 30\degree$ and with no prominent flux emergence nearby. Thereafter, we look for the signature of emergence for these BMRs in the magnetogram by going back in time before they are detected by AutoTAB. Snapshots of 12 such BMRs and their evolutions are shown in 
\Fig{fig:app2} along with the evolution of $\Phi$, $\gamma$, and $D$ in \Fig{fig:app_tilt}.
Our observations based on these selected BMRs reveal that 8 of them emerge with definite tilt, while the remaining emerge with nearly zero tilt. Furthermore, we also note that the BMR signatures emerging with or without tilt do not depend on the latitude of their emergence, the time it takes to be detected by AutoTAB, or their flux strength. The factors determining a BMR emerge with or without tilt remain unclear and require further exploration. Nevertheless, it is noted that nearby flux emergence can influence the tilt of the emerging BMRs.

In summary, as observed in case studies, and based on \Fig{fig:fig3}, we confirm that Joy's law trend is clearly evident at the first detection of tracking. This suggests that some tilt is imparted to the BMRs below the surface before they are observed in the magnetograms, and 
the major cause of the BMR tilt can be the Coriolis force. 

% In the next section, we explore the flux dependence on the tilt of the BMRs.

%In accordance with the thin-flux tube model, the Coriolis force strongly influences flux tubes near the BCZ, imparting an initial tilt as they ascend. However, turbulent convection in the photosphere affects rising flux tubes, particularly weaker BMRs, leading to a loss of the initial tilt from the Coriolis force and affecting the tilt observed at the photosphere. Stronger BMRs, more resistant to turbulence, may retain the initial tilt. This effect of turbulence is discussed in detail in \Sec{sec: JL}. The initial tilt observed in the BMR

%\subsection{Tilt dependence on the flux}
\subsubsection{Flux dependence of tilt}
\label{sec:flux}

\begin{figure*}[t]
\centering
\includegraphics[width=\textwidth]{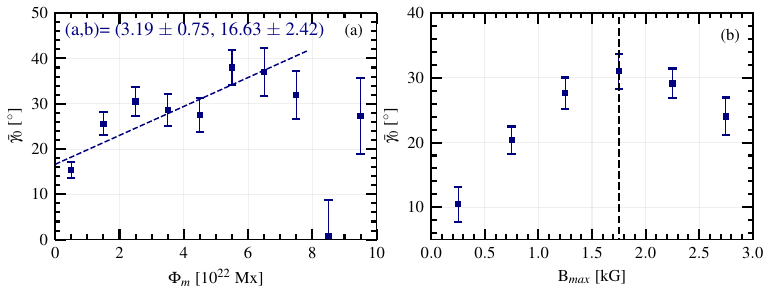}
\caption{(a) Mean $\gamma_0$ is calculated in each flux bin ($\Phi_m$) of length $1 \times 10 ^{22}$~Mx and is plotted as a function of $\Phi_m$. The dashed blue line represents the straight line fit ($\gamma_0 = a \Phi_m + b$), excluding the last two points with parameters mentioned on top of the panel.
Here, the error bars represent the standard error in each flux bin, which is bigger in the high flux bins because of less number of BMRs.
(b) Mean $\gamma_0$ is calculated in each \bmax\ bin of length 0.5~kG and is plotted against \bmax. Dashed black vertical line represents \bmax\ at 1.75~kG.
}
\label{fig:fig5b}
\end{figure*}

As the BMR flux is observed to show a wide variation in magnitude (more than three orders), we expect to detect a variation of tilt with the magnetic flux, which was one of the predictions of the thin flux tube model as discussed in \Sec{sec:expect}.  
To assess the tilt dependence on flux, we compute  $ \overline{\gamma_0} = \langle \gamma \rangle /\langle\sin{ \lambda} \rangle$
in each $10^{22}$~Mx flux bin and plot them against the $\Phi_m$ in \Fig{fig:fig5b}(a).
We note that in computing  $\overline{\gamma_0}$, a normalization factor  $\langle \sin\lambda \rangle$ is used to get rid of the 
%\sin\lambda$ factor  appearing 
latitudinal dependency in Joy's law. 
However, this quantity, $\overline{\gamma_0}$, is strictly not the slope of Joy's law ($\gamma_0$).  Taking the average on both sides of \Eq{eq:1}, we can regard this factor $\overline{\gamma_0}$ as the mean slope of Joy's law. Instead of computing it in this way, if we compute $\gamma_0$ in the traditional method (i.e., by fitting tilt v/s latitude variation) in each flux bin, then we get a statistically insignificant value of $\gamma_0$ due to limited data in some bins (however see the next section).
Despite this, we observe a consistent increase in $\overline{\gamma_0} $ with $\Phi_m$ until $7 \times 10^{22}$~Mx, and the best fit is found to be linear ($\overline{\gamma_{0}} = a \Phi_m + b$).

This dependence is much stronger than the one predicted by the thin flux tube model of \citet{FF1994}; see \Eq{eq:1}. However, we note that the measured total unsigned flux of the BMR might be an averaged representation of flux inside the tube, and a precise simulation-like behavior cannot be anticipated from observational data. Nevertheless, any observed dependence of $\overline{\gamma_0}$ on $\Phi_m$ indicates a significant dependence of tilt on the flux of the BMR and thus supports the thin flux tube model.

After $7 \times 10^{22}$~Mx, we observe an indication of a decrease of $\overline{\gamma_0}$ with $\Phi_m$
%, and \hl{the reasons for this behavior remain uncertain} 
(assuming that there is no saturation limit in the measurements). The reliability of these data points in the graph is 
%also 
however 
compromised due to the limited number of high-flux BMRs. Nevertheless, this decrease could be due to the dominance effect of the magnetic field at a large flux (as \bmax\ increases with $\Phi_m$; \Fig{fig:fig5a}). While the tilt increases with the magnetic flux, it decreases with the field in the flux tube; see \Eq{eq:1}. In high flux regime, the effect of magnetic field dominates over the flux and it causes to decrease the tilt. 
This decrease of tilt is referred to as the tilt quenching by \citet{JK2020} \citep[also see the inclusion of tilt quenching in dynamo models][]{KM17,KM18}, who observed a decrease in $\gamma_0$ with \bmax\ \citep[see Figure 4(a) in][]{JK2020}. 
%Further examination of the relationship between \bmax\ and $\overline{\gamma_0}$ using tracked BMR data is presented in \Fig{fig:fig5b}(b), 
In our  tracked BMR data also we find 
%revealing 
a similar trend between \bmax\ and $\overline{\gamma_0}$ as observed by \citet{JK2020}—an initial increase of $\gamma_0$ with \bmax\, followed by a mild descend beyond 1.75~kG. However, the reliability of the data points for higher \bmax\ bins is limited for the same reasons, as highlighted in \Sec{flux_bmax} (saturation limit in measurements).

The trends in \Fig{fig:fig5b} indicate a dynamic interplay of forces on rising flux tubes. Small and moderate field and flux might represent a region where the drag and Coriolis forces dominate over the magnetic buoyancy, while the regime with \bmax$ > 1.75$~kG (and $\Phi_m \gtrsim 7 \times 10^{22}$) represents the magnetic buoyancy dominated regime, resulting in reduced tilt.

%\subsection{A Revisit to Joy's law: Flux dependence}
\subsubsection{Flux dependence of Joy's law}
\label{JL}

%We have seen from previous sections that analysis of tracked information of BMRs from the AutoTAB catalog supports the thin flux tube model as a viable explanation for the formation of BMRs. With this in mind, this section delves into the assessment of the validity of Joy's law in the framework of the thin flux tube model.
%In \Sec{in_jl}, we have shown that  the average tilts and latitudes of BMRs, computed during the times when their flux exceeds 80\% of peaks, follow Joy's law. Now shall check the robustness of this law at different flux bins. 
%\citet{DC1993} predicts without any further constraints invoked, BMRs emerge with an expected tilt with appropriate latitudinal dependency due to the effect of the Coriolis force. From the observational perspective, tilt data is extremely noisy, and Joy's law trend is obvious after averaging. \citet[]{LC2002, WF2013} suggests that the major contribution to the scatter in the tilt is due to the turbulent convection affecting the flux tube as it rises through the CZ, especially on lower flux regions. Evidence of the same is seen in \Fig{fig:flux_evol}(b).
%BBK: Not sure that figure suggest the evidence of the scatter.  

\begin{figure}[t]
\centering
\includegraphics[width=0.5\textwidth]{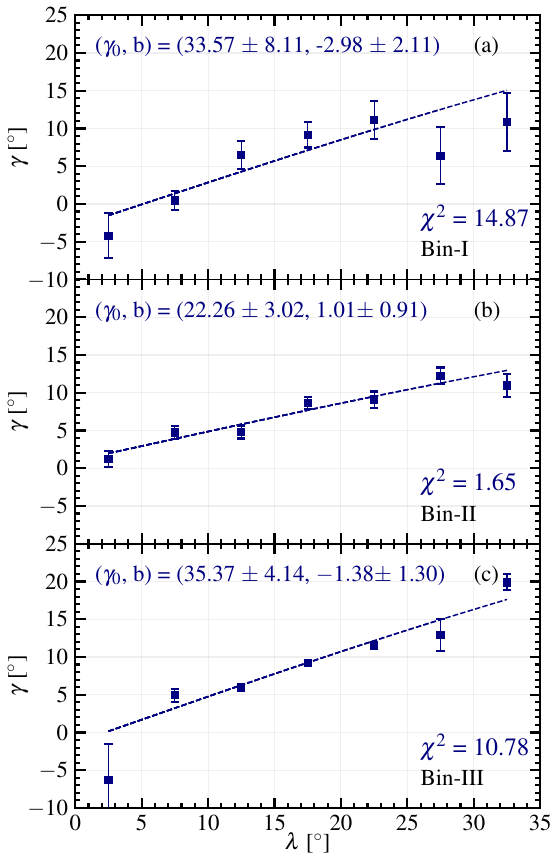}
\caption{\abs{Joy's law dependence for BMRs with flux %(a) Bin-I ($10^{20}$ -- $10^{21}$~Mx), (b) Bin-II  ($10^{21}$ -- $10^{22}$~Mx), and (c) Bin-III ($10^{22}$ -- $10^{23}$~Mx).
(a) Bin-I ($1.72\times10^{19}$ -- $3.13\times10^{21}$\,Mx), (b) Bin-II  ($3.13\times10^{21}$ -- $2.38\times10^{22}$\,Mx), and (c) Bin-III ($2.38\times10^{22}$ -- $9.92\times10^{24}$\,Mx).}
The dashed blue line represents Joy's law fit $\gamma = \gamma_0 \sin \lambda + b$ with fitting parameters mentioned on the panel in blue.}
\label{fig:fig6}
\end{figure}

To further explore the magnetic flux dependence of tilt, here we revisit Joy's law. 
We have seen in \Fig{fig:fig1} that the tracked BMRs from the AutoTAB catalog collectively obey Joy's law.  However, we have already noted that the tilt data is extremely noisy (e.g., \Fig{fig:flux_evol}(b)), and Joy's law trend is evident only after averaging.  A possible contribution to the scatter in the tilt is due to the turbulent convection affecting the flux tube as it rises through the CZ \citep[e.g,][]{LC2002, WF2013}, and thus, the effect of scatter is more prominent in lower flux bins.  Already, we have seen some evidence of it in \Fig{fig:flux_evol}(b).
We, therefore, segregate the BMRs into three bins with an equal number of data points in each bin.  To keep the same number of data points in each bin, the flux ranges in these bins become: \abs{$1.72\times10^{19}$ -- $3.13\times10^{21}$\,Mx (Bin-I), $3.13\times10^{21}$ -- $2.38\times10^{22}$\,Mx (Bin-II), and $2.38\times10^{22}$ -- $9.92\times10^{24}$\,Mx (Bin-III).} %abs{} 
We note that although the maximum flux of BMR in Bin III goes to a very large value, there are only a few BMRs above $2\times10^{23}$ Mx; see \Fig{fig:fig5a} (in fact, the median flux values in each bin are $6.64\times10^{20}$~Mx, $9.81\times10^{21}$~Mx, and $5.28\times10^{22}$~Mx, respectively.)
%abs{} 
\Fig{fig:fig6} displays tilt as a function of latitude, along with Joy's law fit, for the BMRs in these bins. We observe that as we move to the higher flux regimes (Bins-II and III), the scatter decreases, which agrees with the theoretical expectation that the stronger BMRs are less buffeted by convection.

Also, seeing a large scatter around Joy's law and a large error in the fitted parameters in Bin-I, having BMR flux $< 3.13 \times 10^{21}$~Mx, we raise a doubt whether Joy's law behavior is valid only for stronger BMRs.  Interestingly, if we discard this bin, then from panels (b) and (c), we  observe that the slope of Joy's law $\gamma_0$ increases as we move from Bin-II (having mean flux $1.11\times 10^{22}$~Mx) to Bin-III (mean flux $4.93 \times 10^{23}$~Mx).  This $\gamma_0$ increase is in agreement with the theoretical prediction and, in particular, with the result presented in the previous section (\Sec{sec:flux}) that $ \overline{\gamma_0} = \langle {\rm \gamma} \rangle /\langle {\rm \sin \lambda } \rangle$ increases with flux.

%\abs{}
We note that in \Fig{fig:fig6}, the result also remains consistent if we include only the BMRs within the flux range of $10^{20}$ -- $10^{23}$~Mx (i.e., if we cut the two tails of the flux distributions) and 
bin the data in equal numbers. 
 Interestingly, if we bin the data in equal flux bin such that the flux ranges become $10^{20}$ -- $10^{21}$~Mx (Bin-I), $10^{21}$ -- $10^{22}$~Mx (Bin-II), and $10^{22}$ -- $10^{23}$ Mx (Bin-III), then the value of $\gamma_0$ in Bin-II and Bin-III are comparable.  However, the decrease in scatter from Bin I to Bin II remains consistent in this case as well. 
This suggests that Joy's law fitting is sensitive to how we bin the data points. 
This could be why previous authors could not find a systematic increase of $\gamma_0$ with the flux \citep{SK2008, SK2012}.
%\abs{}
Therefore,  $\langle {\rm \gamma} \rangle /\langle {\rm \sin \lambda } \rangle$ may be a better quantity when measuring the flux dependent of BMR tilt as done in \Sec{sec:flux}.

%Although we observe that scatter around individual points decreases as we go to higher flux bins, the flux dependence on Joy's law slope is absent. The Joy's law plots of all three bins are fitted with Joy's law function, and the fitting parameters, along with uncertainties, are mentioned in the panels of \Fig{fig:fig6}. The $\gamma_0$ of all three bins overlap, failing to show the dependence of flux. The similar behavior was reported by \citet{SK2008, SK2012}. This result also agrees with simulation studies by \citet{WF2013} within similar flux bins.
%BBK: Do we need the text below?
%Upon considering $\chi^2$ values for each fit, the sinusoidal fit or the straight-line fit conventionally used in related studies may be appropriate only for the higher flux regions.  
%Upon considering $\chi^2$ values for each fit, we find that, for the higher flux bin (Bin-II, Bin-III), sinusoidal fit and straight line fit have $p$--values of 0.9, indicating a rather good fit. Both the fitting functions fail in Bin-III, and BMRs with lower flux values show a $p$--value of only 0.39. 

%\subsection{Tilt scatter: Dependence on separation}
\subsubsection{Variation of tilt fluctuations with footpoint separation}
\label{sec:scatter}

\begin{figure}[t]
\centering
\includegraphics[width=0.5\textwidth]{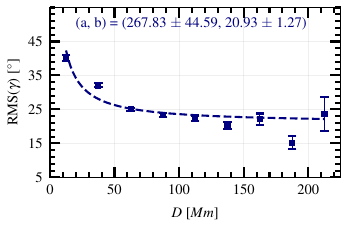}
\caption{Mean root-mean-square of the tilt (RMS($\gamma$)) calculated in each footpoint separation ($D$) bin of length 25~Mm and is plotted as a function of $D$.}
\label{fig:fig7}
\end{figure}

Based on the thin flux tube model, it is expected that as the tubes rise through CZ, they are buffeted by convection. Including turbulence in the numerical model of thin flux tube, \citet{LF1995} predicted dependence of the tilt angle fluctuations i.e., root-mean-square value of the tilt RMS($\gamma$) with the footpoint separation $D$. They showed that as $D$ increases, (i) the averaging effect of small-scale wiggles over the flux tube is greater, leading to straight rise of the tube and (ii) the flux increases, which consequently decreases the rise time or the time for the interaction with convection. They showed that RMS($\gamma$) scales as $1/D$. 
% The simulation studies by \citet{LF1995} show that the tilt angle fluctuations arise in the model from hydrodynamic turbulence as the $\Omega$ loop migrates through the CZ. The effect of the same will be visible in the observation as the rms value of the tilt (RMS($\gamma$)) will decrease as the polarity separation ($D$) increases. 
%Further, from the white-light observation  \citet{FF1994} showed that 
%\({\rm RMS}(\gamma) \simeq 10\degree (D/100)^{-3/4}\). We investigate this behavior with the AutoTAB catalog in \Fig{fig:fig7}. \textcolor{red}{Further to be discussed with Bibhuti.} 
We investigate this behavior by showing the variation of RMS($\gamma$) with $D$ from the AutoTAB catalog in \Fig{fig:fig7}. 
We observe that ${\rm RMS}(\gamma)$ decreases with increasing footpoint separation exactly reported by \citet[${\rm RMS}(\gamma) = a/D + b$]{LF1995}.
%with parameters mentioned in the panel. 
%The plot also depicts ${\rm RMS} (\gamma) \propto  D^{-3/4}$ dependence as red dashed line. 
We note that from white-light observations,  \citet{FF1995} also found a similar dependence: 
\({\rm RMS}(\gamma) \simeq 10\degree (D/100)^{-3/4}\).
%\({\rm RMS}(\gamma) \simeq 10\degree (D/100)^{-3/4}\).
%We observe that both the relations capture the trend in the data, but (${\rm RMS} (\gamma) = a/D + b$) emerges as the better fit compared to the relation $\Delta\gamma \propto  D^{-3/4}$.

\section{Conclusions} \label{sec:conclusion}

The thin-flux tube model provides the simplest explanation for the formation of BMRs, postulating magnetically buoyant tubes anchored at the convection zone. According to this model, the Coriolis force imparts tilt to the emerging BMRs. Despite its popularity, the theory lacks observational backing. This study undertakes an investigation into the validity of the model's assumptions using the AutoTAB catalog. 

AutoTAB is an in-house developed algorithm to automatically detect and track the BMRs from MDI (1996--2011) and HMI (2010--2023) LOS magnetogram data. The resulting comprehensive catalog documents the evolution of 12,173 BMRs on the nearside of the Sun. As a sanity test, we first assign a single representative values for BMR properties (\bmax, \bmean, $\Phi_m$, $\gamma$) by averaging their values during the times when their flux exceed 80\% of the peak values. We then show that BMRs follow (i) Joy's law trend ($\gamma_0 = 29.56\degree$), (ii) Gaussian-like tilt distribution ($\mu = 7.78 \pm 0.43$), and (iii) magnetic flux vs field dependence, all are consistent with previous studies. With these basic tests, we proceed with detailed analyses of our tracked BMRs to validate the thin flux tube model as a theory behind the BMR formation. 

%In this study, we analyze the evolving properties of BMR tilts and the related quantities through the AutoTAB catalog to understand the mechanism behind the formation of BMRs. AutoTAB is an in-house developed algorithm to automatically detect and track the BMRs from MDI (1996--2011) and HMI (2010--2023) LOS magnetogram data. The resulting comprehensive catalog documents the evolution of 12,173 BMRs on the nearside of the Sun. 
%A single representative values for properties (\bmax, \bmean, $\Phi_m$, $\gamma$) are assigned based on their flux peak instances for analysis. The assigned values collectively demonstrate a consistent Joy's law trend ($\gamma_0 = 30.22\degree$) and a characteristic tilt distribution ($\mu = 7.94 \pm 0.43$), aligning closely with previous studies. The correlation observed between $\Phi$ and \bmax\ values suggests that higher $\Phi_m$ regions correspond to stronger \bmax\ values, with instrument saturation influencing magnetic field measurements in higher flux regions.

%\begin{itemize}
    %\item 
    The buoyant rise of flux tube assumption is validated by examining footpoint separation ($D$) over the BMR's tracked lifetime. Our findings indicate a rapid increase in $D$ during the initial phase of BMR evolutions, followed by a gradual rise and eventual saturation towards the end of their lifetime. 
    The rapid increase is primarily attributed to the longitude separation ($\Delta\phi$), particularly pronounced in higher flux regions. This may imply a connection between rapid $D$ growth and footpoint tethering from the convection zone, where the immediate effect of the same manifests as a sudden increase in $\Delta\phi$.
   
    %\item 
    In line with the thin-flux tube model, 
    BMRs are expected to appear with tilt
    at the onset of the emergent phase due to the effect of Coriolis force 
    during the rise of flux tubes. Our analysis of Joy's law trend in tracked BMRs of LT class 
    supports this expectation by demonstrating a clear Joy's law trend in their first detection. This was further evaluated through case studies of individual BMRs selected from different phases of cycles and strengths. 
    Our findings reveal a nuanced scenario, where the signatures of some BMRs 
    emerge with zero tilt and develop it at later phases,  while the rest exhibit a significant tilt from the beginning.   
    This tilt behavior is independent of the emergence latitude or cycle phase, indicating a potential contribution of Coriolis force to a part of observed tilt in BMRs.
    
    %\item 
    Further, based on the thin flux tube model, we expect the tilt to increase with the increase of magnetic flux \citep{DC1993, FF1994}; also see \Eq{eq:1}. 
    We explored this flux dependence of tilt 
    by %normalizing it with $\sin(\lambda)$, obtaining $\overline{\gamma_0}$. 
    computing  $\overline{\gamma_0} =  \langle \gamma \rangle / \langle \sin \lambda \rangle$. 
    Our results reveal a linear increase in $\overline{\gamma_0}$ with $\Phi_m$, signifying a pronounced tilt dependence on magnetic flux. A similar trend is observed concerning \bmax, with an initial increase in $\overline{\gamma_0}$ for lower \bmax\ regions, followed by a decrease after 1.75~kG. %Nevertheless, our observations suggest a dynamic interplay between forces influencing rising flux tubes, with drag and Coriolis forces affecting lower flux regions, while higher flux regions exhibit the dominance of magnetic buoyancy, resulting in reduced tilt. 
    %Additionally, we assessed Joy's law trend in different flux bins and failed to find any flux dependence on $\gamma_0$. The observations reveal a significant scatter associated with smaller flux regions, decreasing with higher flux values, indicating a stronger influence of turbulent convection on lower flux regions, in line with prior studies by \citet{LF1995, WF2013}.  
    We further observed that Joy's law variation is flux dependent. The Joy's law is less significant (and has a large scatter around the mean trend) at the small flux bin (below about $10^{21}$~Mx). The slope of Joy's law increases as we move the BMR flux bin from $10^{21}$~Mx to $10^{23}$~Mx. This result is again in agreement with the prediction of the thin flux tube model. 

    %\item 
    Finally, based on the thin flux tube model, we expect that the tilt fluctuation (rms value) to depend on the footpoint separation \citep[e.g,][]{LF1995}.  BMRs having large footpoint separation have less chance to be buffeted by the convection. From our data, we find that the tilt fluctuation decreases inversely with the footpoint separation as predicted by the model of \citet{LF1995}.

Although our analysis provides some support to the thin flux tube model and hints at Coriolis force as the reason behind a part of the tilt observed in the BMR, further study using richer data is needed to strengthen our conclusion. Notably, we have to carefully identify the emergent phase of a BMR and automatically compute the tilt of a large number of BMRs to check the statistical reliability of tilt at the very early phase of BMR. We also need to explore the flux dependence of the tilt angle using a longer dataset.  
    
    %Although our analysis provides some support to the thin flux tube model and hints Coriolis force as the reason behind a part of the tilt observed in the BMR, further study is needed to stablish the ro there is a disagreement upon what defines the emergent phase of the BMR through the literature. Further detailed analysis, possibly with a larger dataset, is required to investigate if the BMR emerges with or without tilt and the dependence between tilt and flux of the BMRs.

\begin{acknowledgments}
A.S. sincerely expresses her gratitude to ARIES, Nainital, for the warm hospitality extended during the project’s initial phase. 
B.B.K. acknowledges the ﬁnancial support from the Department of Science and Technology (SERB/DST), India, through the Ramanujan Fellowship (project No. SB/S2/RJN-017/2018). 
A.S. and B.B.K. thank Sami Solanki, Natalie Krivova, and Robert Cameron for their valuable comments and discussions during their visit to MPS, which helped in improving the manuscript.
The observational data, the LOS magnetograms from MDI and HMI, used in this article were obtained via
JSOC, courtesy of the NASA/SOHO and NASA/SDO science teams.
\end{acknowledgments}

\appendix

\section{Examples of evolutions of BMRs}

Panels (a)-(f) and (g)-(l) in \Fig{fig:app2} 
present snapshots to illustrate the evolution of 12 distinct BMRs tracked by AutoTAB from Solar Cycle 23, starting from the point when the BMR signatures first emerged.  The snapshots with red rectangles are produced from the times when AutoTAB could track the BMRs, while the snapshots with blue rectangles show their evolutions back in time. 
The criteria for region selection were as follows: (i) Individual BMRs' emerging signatures should fall within $\pm 55\degree$ longitude and $\pm 30\degree$ latitude, and (ii) There should be no significant flux emergence in the vicinity. The ``T" value at the bottom indicates the evolutionary stage of the BMR. ``T:0'' signifies the initial detection by AutoTAB. Pre- and post-detection times are indicated in hours, denoted by negative and positive signs. The time and date of first detection is mentioned at the top of the panel of ``T:0". Bracketed numbers denote the mean heliographic latitude and longitude for each evolutionary stage.

Among the selected regions, we observed that eight BMRs emerged with a significant tilt at the initial phase itself, 
while others emerged with nearly zero tilt. The evolution of $\gamma$, $\Phi_m$, and $D$ during their tracked lifetime for all the selected BMRs are illustrated in panels (a)-(l) in \Fig{fig:app_tilt}.

\begin{figure*}[!htbp]
    \gridline{\fig{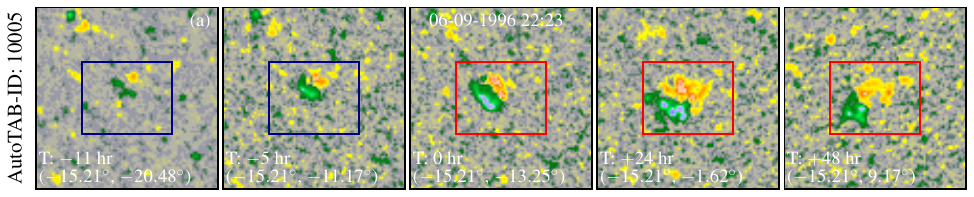}{\textwidth}{}}
    \vspace{-10mm}
    \gridline{\fig{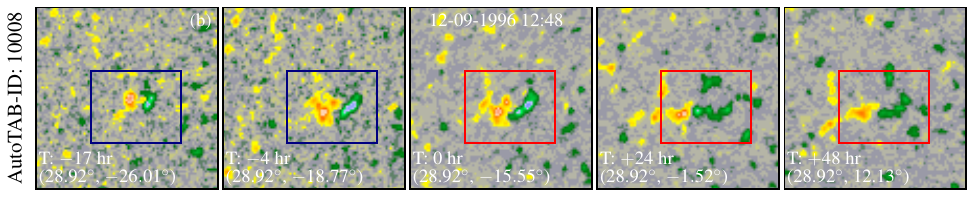}{\textwidth}{}}
    \vspace{-10mm}
    \gridline{\fig{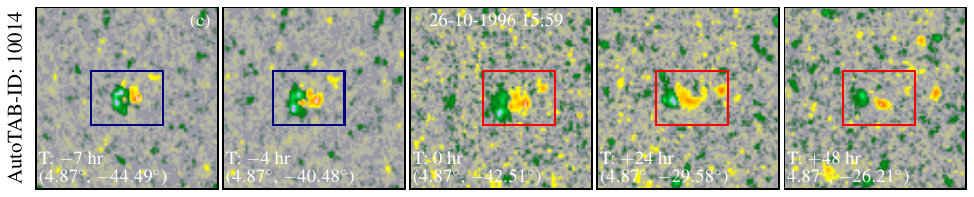}{\textwidth}{}}
    \vspace{-10mm}
    \gridline{\fig{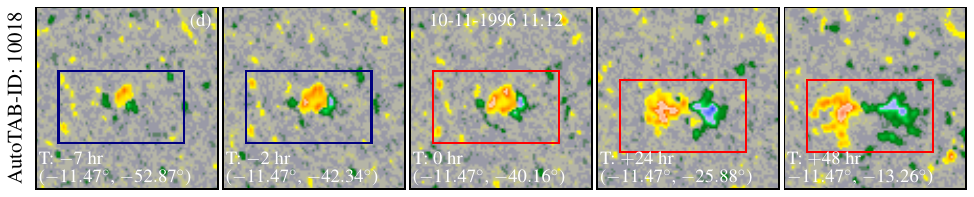}{\textwidth}{}}
    \vspace{-10mm}
    \gridline{\fig{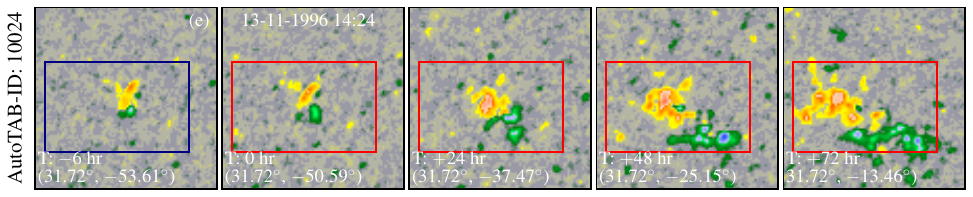}{\textwidth}{}}
    \vspace{-10mm}
    \gridline{\fig{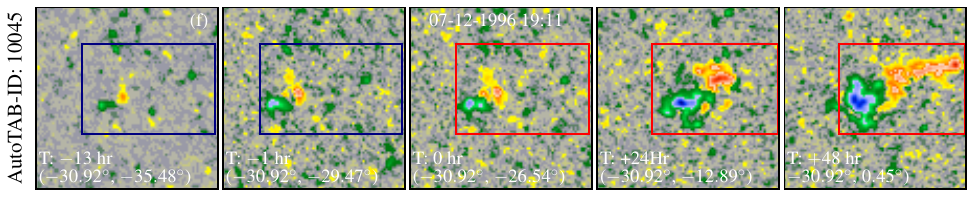}{\textwidth}{}}
    %\label{fig:app1}
\end{figure*}

\begin{figure*}[!htbp]
    \gridline{\fig{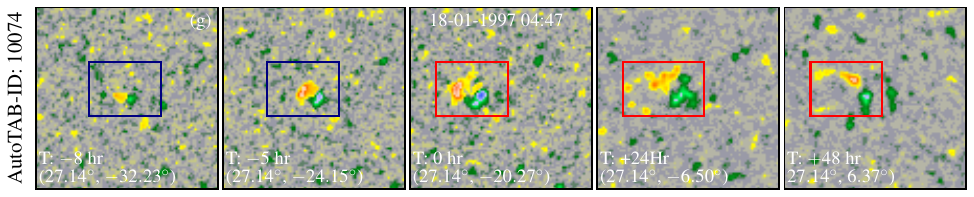}{\textwidth}{}}
    \vspace{-10mm}
    \gridline{\fig{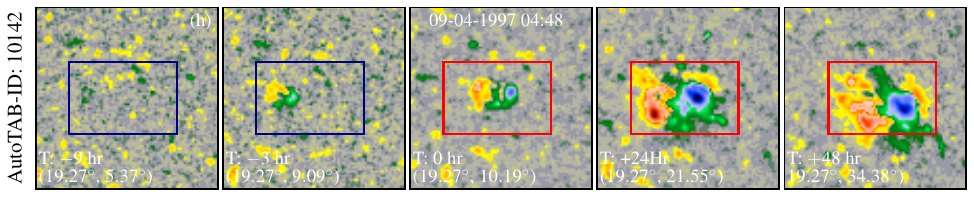}{\textwidth}{}}
    \vspace{-10mm}
    \gridline{\fig{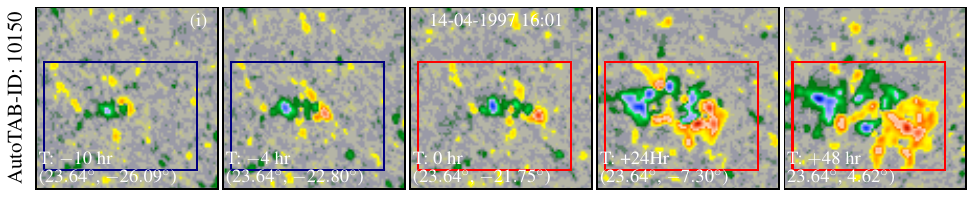}{\textwidth}{}}
    \vspace{-10mm}
    \gridline{\fig{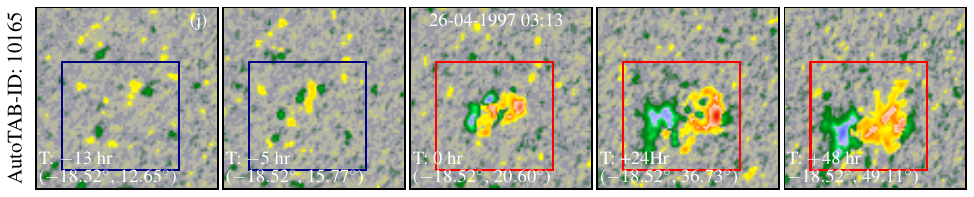}{\textwidth}{}}
    \vspace{-10mm}
    \gridline{\fig{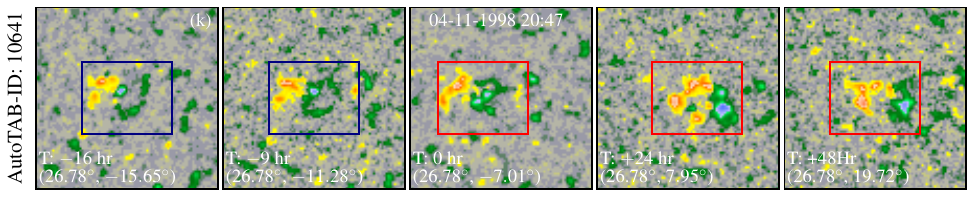}{\textwidth}{}}
    \vspace{-10mm}
    \gridline{\fig{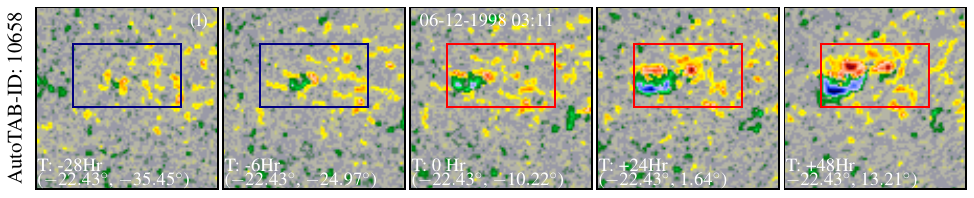}{\textwidth}{}}
    \caption{Snapshots of the evolution of 12 BMRs tracked by AutoTAB with 1~day cadence. See text for details. Numbers in the brackets at the bottom of panels represent the mean latitude and longitude of polarities of the BMR.}
    \label{fig:app2}
\end{figure*}

%\clearpage

\begin{figure*}[!htbp]
    \gridline{\fig{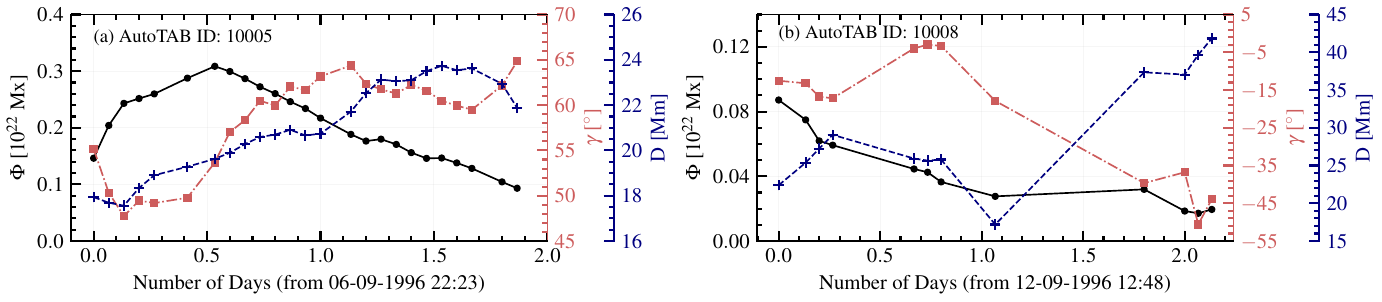}{0.98\textwidth}{}}
    \vspace{-10mm}
    \gridline{\fig{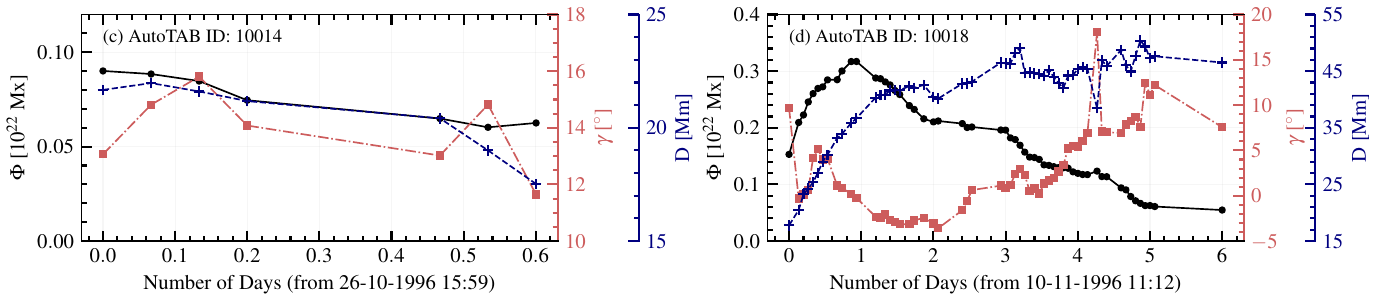}{0.98\textwidth}{}}
    \vspace{-10mm}
    \gridline{\fig{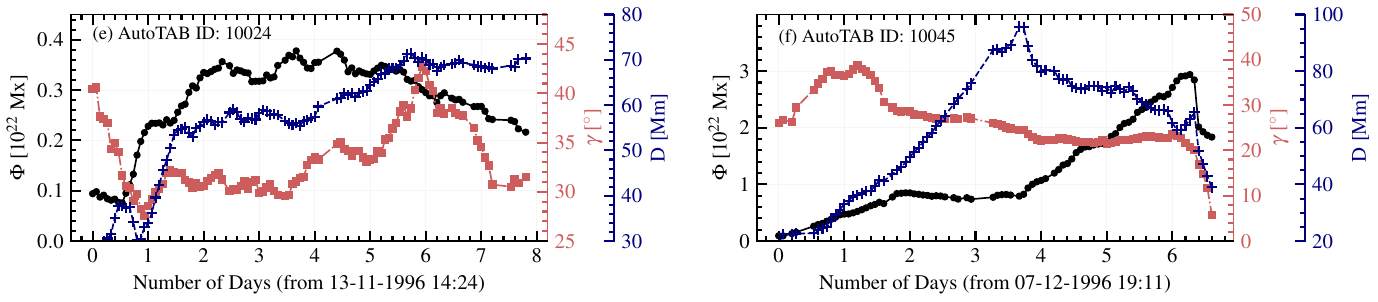}{0.98\textwidth}{}}
    \vspace{-10mm}
    \gridline{\fig{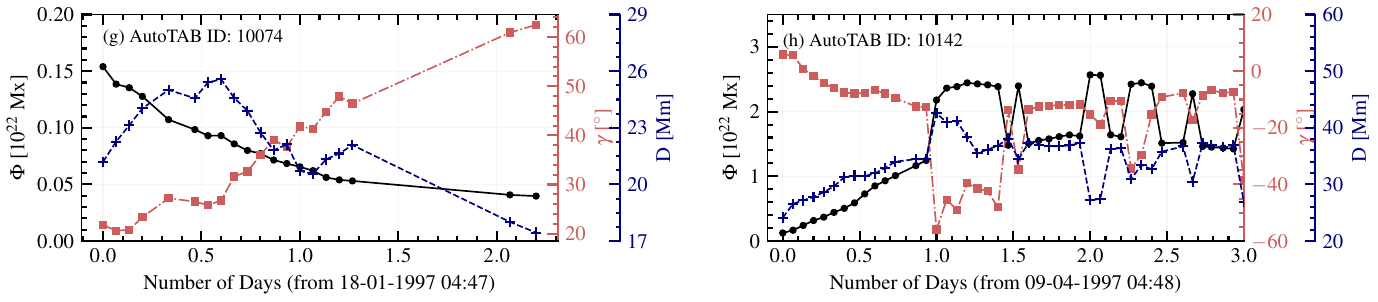}{0.98\textwidth}{}}
    \vspace{-10mm}
    \gridline{\fig{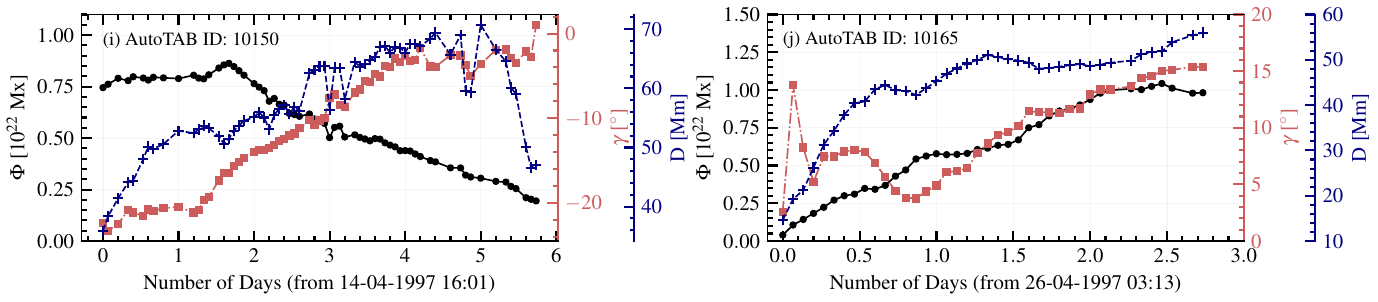}{0.98\textwidth}{}}
    \vspace{-10mm}
    \gridline{\fig{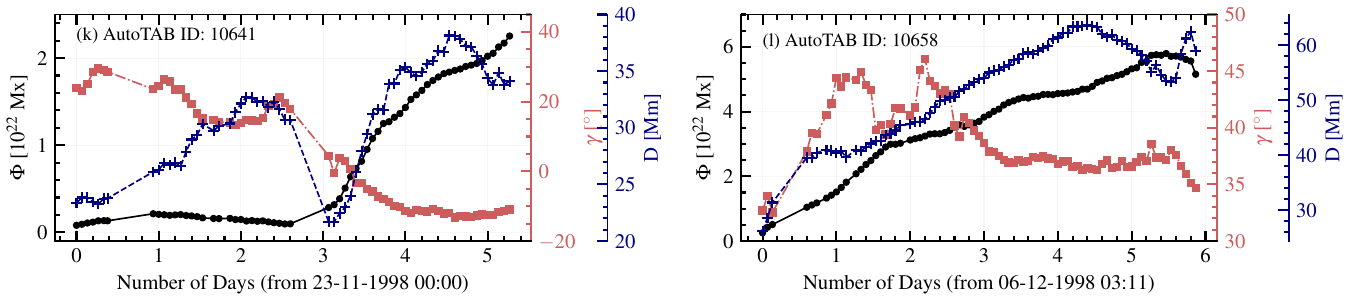}{0.98\textwidth}{}}
    \vspace{-10mm}
    \caption{Time evolutions of flux, tilt, and footpoint separation of 12 BMRs whose evolution are shown in \Fig{fig:app2}.}
    \label{fig:app_tilt}
\end{figure*}
\clearpage

%\bibliography{references}{}
%\bibliographystyle{aasjournal}

\end{document}